\documentstyle[twoside,fleqn,espcrc2,epsf]{article}
\title{Can FCNC transition $c\to ul^+l^-$ be seen in $D\to Vl^+l^-$ decays?}
\author{S. Prelov\v sek\address{J. Stefan Institute, Jamova 39, P. O. Box 3000, 1001 
Ljubljana, 
Slovenia}, S. Fajfer$^{\rm a}$ and  P. Singer\address{ Department of Physics, Technion - Israel Institute  of Technology, 
Haifa 32000, Israel}}
\begin{document}
\begin{abstract}  
The decays $D\to Vl^+l^-$ present in principle the 
opportunity to observe the 
short distance FCNC transition $c\to ul^+l^-$, which is sensitive 
to the physics 
beyond the Standard Model. 
We analyze the $D \to V l^+ l^- $ 
decays within the Standard Model, where in addition to the short distance 
dynamics also the long distance dynamics is present.  The short distance contribution due to $c\to ul^+l^-$ transition, which is present only in the Cabibbo suppressed decays,  is found to be three orders of magnitude smaller than  the long distance contribution. The branching ratios well above $10^{-7}$ for  Cabibbo suppressed decays could signal new physics. The most frequent 
decays   are the Cabibbo allowed decays 
$D_s^+\to\rho^+\mu^+\mu^-$ and $D^0\to\bar K^{*0}\mu^+\mu^-$, which  are expected at the branching ratios of $3\cdot 10^{-5}$ and $1.7\cdot 10^{-6}$, respectively.  These rates are not much lower than the present experimental upper limit. 
\end{abstract}
\maketitle

In the charm sector the  phenomena like $D^0-\bar D^0$ mixing, 
CP-violation and  rare decay probabilities are small, which makes them good 
candidates as probes for New physics with small background from the
Standard Model. 
The smallness of the short-distance (SD) $c\to 
ul^+l^-$ rate within the Standard Model suggests that the decays $D\to Vl^+l^-$ ($V$ is light vector meson) could serve  
as a possible 
good window to non-standard contributions to the flavour-changing neutral 
transition $c\to ul^+l^-$ \cite{SCHWARTZ,beyond}. 
We analyse all $D \to V l^+ l^- $ 
decays within the Standard Model, where except for the short distance 
dynamics also the long distance (LD) dynamics is present.  We calculate the SD and LD contributions using the hybrid model, which combines heavy quark and chiral perturbation theory.
 The SD contribution due to $c\to ul^+l^-$ is present only in the 
Cabibbo suppressed decays $D^0 \to \rho^0 l^+ l^-$, $D^0 \to \omega l^+ l^-$ 
$D^0 \to \phi l^+ l^-$, $D^+ \to \rho^+ l^+ l^-$ and 
$D_s^+ \to K^{*+} l^+ l^-$. In this decays our motivation is to determine the relative magnitude of SD and LD contribution. Our results should provide the appropriate theoretical 
background against which possible signals of new physics are searched for.     Motivated by the experimental searches, we analyze also the  Cabibbo allowed decays 
($D^0\to K^{*0}l^+l^-$ and $D_s^+\to \rho^+l^+l^-$), which are the best candidates for their early detection, and the doubly Cabibbo 
suppressed decays ($D^+\to K^{*+}l^+l^-$ and $D^0\to K^{*0}l^+l^-$). Here  the 
signals from new physics are not expected from the theoretical models usually 
considered. 

On the experimental side, so far there are only upper bounds on decays $D\to 
Vl^+l^-$ from E653  and CLEO \cite{exp}, in the range 
$10^{-3}-10^{-4}$, but these are expected to improve in the future.

This work has been presented in detail in \cite{vll}.

\vspace{1.5mm}

The {\bf long distance} contribution in $D\to Vl^+l^-$ decays is due to the effective nonleptonic weak Lagrangian, which induces the weak transition between the initial and final hadronic state. The weak transition has to be accompanied by the emission of a virtual photon, which finally decays  decays into a lepton antilepton pair.
The  effective nonleptonic weak Lagrangian responsible for charm meson decays is
\begin{eqnarray}
\label{weak}
{\cal L}_{LD}&= & -{G_F \over \sqrt{2}} V_{uq_i}V^*_{cq_j}   
~[ a_1 ({\bar u} q_i)^{\mu}
({\bar q_j}c )_{\mu} \nonumber\\
& +&  a_2 ({\bar u} c)^{\mu} ({\bar q}_j q_i)_{\mu}]~, 
\end{eqnarray}
where $({\bar \psi}_1\psi_2)^\mu\equiv
{\bar \psi}_1\gamma^\mu(1-\gamma^5)\psi_2$ and  $q_{i,j}$ represent the fields of $d$ or $s$ quarks. 
There are two kinds of LD contribution:  in the {\it resonant} mechanism, apart from the final vector meson $V$, an additional neutral vector meson  $V_0$ is produced, which converts to a photon 
through vector meson dominance ($D\to VV_0\to V\gamma \to Vl^+l^-$; $V^0$ is $\rho$, $\omega$ or $\phi$); in the {\it nonresonant} mechanism  the photon is emitted directly from the initial $D$ or final $V$ meson state.  To calculate the amplitudes we use the factorization approximation and the  hybrid model, which combines 
heavy quark effective theory and chiral perturbation theory \cite{vll}. The details of this framework and its previous application is given in \cite{casrep}.   The relevant degrees of freedom are heavy pseudoscalar ($D$) 
and vector ($D^*$) meson fields  and light pseudoscalar ($P$) and vector ($V$) meson fields. 
Within this approach, the diagrams that contribute to the process under study are given in Fig. 1. The diagrams are divided in three different types according to the factorization: Figs. 1a and 1b represent two types of spectator 
contributions, while Fig. 1c represents the weak annihilation contribution. The square in each diagram denotes the weak transition due to the effective Lagrangian ${\cal 
L}_{LD}$ (\ref{weak}). This Lagrangian contains a product of two left handed 
quark currents $(\bar q_kq_l)^{\mu}$, each denoted by a dot. 
The diagrams $III$ and $IV$ represent nonresonant 
contribution, while all the remaining diagrams represent the resonant contribution. The vertices are evaluated using the hybrid model \cite{vll,casrep}. 

\vspace{1.5mm}

 The {\bf short distance} contribution  due to FCNC $c\to ul^+l^-$ is present only in the Cabibbo suppressed $D\to Vl^+l^-$ decays.  The effective Lagrangian for $c \to u l^+ l^-$ arises from $WW$ exchange box 
diagrams and $Z$ and $\gamma^*$ penguin operators \cite{SCHWARTZ,vll}. The main contribution comes from the intermediate $d$ and $s$ quarks exchange and by using $m_d^2\ll m_W^2$ and $m_u^2\ll m_W^2$ one obtains
\begin{eqnarray}
{\cal L}_{SD}= \frac{G_F}{{\sqrt 2}} \frac{e^2\sum_{i=d,s,b}V_{ci}^*V_{ui} A_i }{8 \pi^2 sin^2 \theta_W } ~
(\bar u c)^{\mu}~\bar l \gamma_{\mu}l~,\nonumber
 \end{eqnarray}   
where the Willson coefficient $A_i$ is given in \cite{vll}.  This gives the branching ratio for inclusive $c\to ul^+l^-$ process 
$\Gamma(c\to ul^+l^-)/ \Gamma(D^0)=2.9~10^{-9}$, while for the exclusive processes $D\to Vl^+l^-$ we evaluate  $\langle V|({\bar u} c)^{\mu}| D\rangle$  using the hybrid model. 

\vspace{1.5mm}

The predicted branching ratios \cite{vll} containing LD and SD contributions and experimental upper bounds \cite{exp} are given in Table 1. Apart from the Cabibbo structure, the 
branching ratios depend mainly on whether the initial  state is 
charged or neutral, with bigger branching ratio in the former case. 
 The Cabibbo allowed decays 
$D^0 \to \bar K^{*0} \mu^+ \mu^-$  and  $D_s^+ \to \rho^+ \mu^+ \mu^-$ 
have the best probability for their early detection. Note that their 
predicted branching ratios  are not far below the present experimental upper 
bound. 
In the Cabibbo suppressed decays the branching ratios due to SD contribution are three orders of magnitude smaller than the LD contribution: $9.5 ~10^{-10}$ for $D^0\to \rho^0(\omega) \mu^+\mu^-$, 
$4.8 ~10^{-9}$ for $D^+\to \rho^+ \mu^+\mu^-$, $1.6 ~10^{-9}$ for $D_s^+\to K^{*+} \mu^+\mu^-$ and $0$ for $D^0\to \phi \mu^+\mu^-$.  We show the distributions $(1 /\Gamma_D ) d \Gamma (D\to V \mu^+\mu^-)/ d q^2$ for the most frequent decay $D^+_s\to \rho^+ \mu^+\mu^-$ and for the Cabibbo suppressed decay $D^0\to \rho^0 \mu^+\mu^- $ in Fig. 2, where we separate SD, nonresonant LD and the total branching ratio. The rates are obviously dominated by the resonant LD part.
It is interesting to remark however, that the SD part is comparable to nonresonant LD part  in $D^0 \to \rho^0 (\omega)\mu^+ \mu^-$  decays.

\vspace{1.5mm}

We have analyzed SD and LD contributions to $D\to Vl^+l^-$ decays within the Standard Model. The predicted branching ratios for Cabibbo allowed decays are not much lower than the present experimental upper bound. In Cabibbo suppressed decays, the SD contributions are smaller that LD ones by three orders of magnitude. Still, the new physics could greatly enhance SD rates and  branching ratios well above $10^{-7}$ for  
Cabibbo suppressed decays could signal new physics. As the present experimental upper bounds for Cabibbo suppressed decays are much higher, they provide a large discovery window.  \\

\begin{table}[h]
\begin{center}
\begin{tabular}{|c||c|c||c|}
\hline
 $D\to $ & theory & exp \\
 $V\mu^+\mu^-$ & $Br(LD+SD)$ & $Br$ \\
\hline \hline
  $ D^0 \to {\bar K}^{*0} $ &  $[1.6-1.9]~ 10^{-6}$ & $<1.18~10^{-3}$  \\
\hline
  $ D_s^+ \to \rho^+ $  & $[3.0-3.3] ~10^{-5}$ &   \\
\hline
\hline
 $ D^0 \to \rho^{0}$ & $[3.5-4.7]~ 10^{-7}$ & $<2.3~10^{-4}$\\
\hline
 $ D^0 \to \omega $ & $[3.3-4.5]~ 10^{-7}$  & $<8.3~10^{-4}$\\
\hline
 $ D^0 \to \phi $ & $[6.5-9.0]~ 10^{-8}$ & $<4.1~10^{-4}$\\
\hline
 $ D^+ \to \rho^+ $ & $[1.5-1.8]~ 10^{-6}$ & $<5.6~10^{-4}$ \\
\hline
 $ D_s^+ \to K^{*+ }$ & $[5.0-7.0]~ 10^{-7}$ & $<1.4~10^{-3}$\\
\hline
\hline 
 $ D^+ \to K^{*+} $ & $[3.1-3.7]~ 10^{-8}$ & $<8.5~10^{-4}$\\
\hline
 $ D^0 \to K^{*0} $ & $[4.4-5.1]~ 10^{-9}$ & \\
\hline
\end{tabular}
\caption{ The predicted branching ratios \cite{vll} (second column), which contain LD and SD contributions, and experimental upper bounds \cite{exp} for Cabibbo allowed, suppressed and doubly suppressed $D\to V\mu^+\mu^-$ decays. }
\end{center}
\end{table}

{\bf Figure Captions:}

{\bf Fig. 1:} Skeleton diagrams of various long distance contributions to 
the decay $D \to V l^+l^-$ resulting from ${\cal L}_{LD}$ (\ref{weak}). 

{\bf Fig. 2:}   The predicted differential branching ratios 
$(1 /\Gamma_D ) d \Gamma (D_s^+ \to \rho^+ \mu^+ \mu^-)/ d q^2$ (Fig. 2a) and 
$(1 /\Gamma_D ) d \Gamma (D^0 \to \rho^0 \mu^+ \mu^-)/ d q^2$ (Fig. 2b)  
as a function of $q^2$ ($q^2$ the invariant $\mu^+ \mu^-$ mass). 
The full line represents to the total  branching ratio,  the  
dot-dashed line  represents the  short distance part, while the dashed line 
represents the nonresonant long distance part. 


\begin{thebibliography}{10}
\bibitem{SCHWARTZ} A. J. Schwartz, Mod.Phys.Lett. A8 (1993) 967.
\bibitem{beyond} G.L. Castro et all. hep-ph/9804368; 
 K.S. Babu et all. , Phys. 
Lett. B 205 (1988) 540.
\bibitem{exp}  CLEO coll., Phys.Rev.Lett. 76 (1996) 3065; E653 coll., Phys.Lett. B 345 (1995) 82.
\bibitem{vll} S. Fajfer, S. Prelovsek and P. Siger, hep-ph/9805461.
\bibitem{casrep} R. Casalbuoni et all., Phys.Rep. 281 (1997) 145.  
\end{thebibliography}
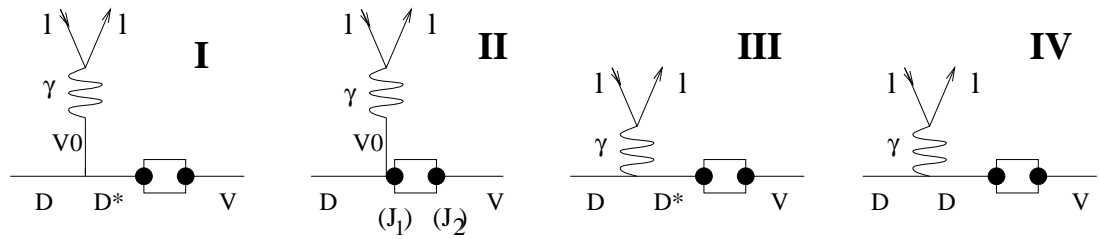
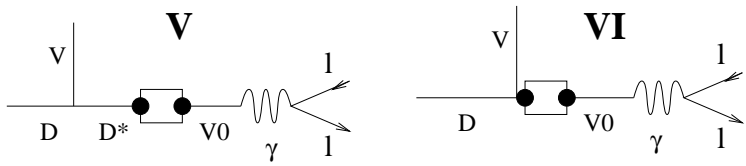
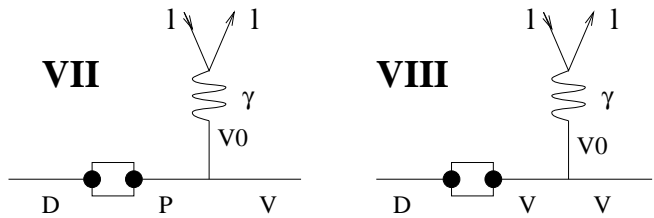
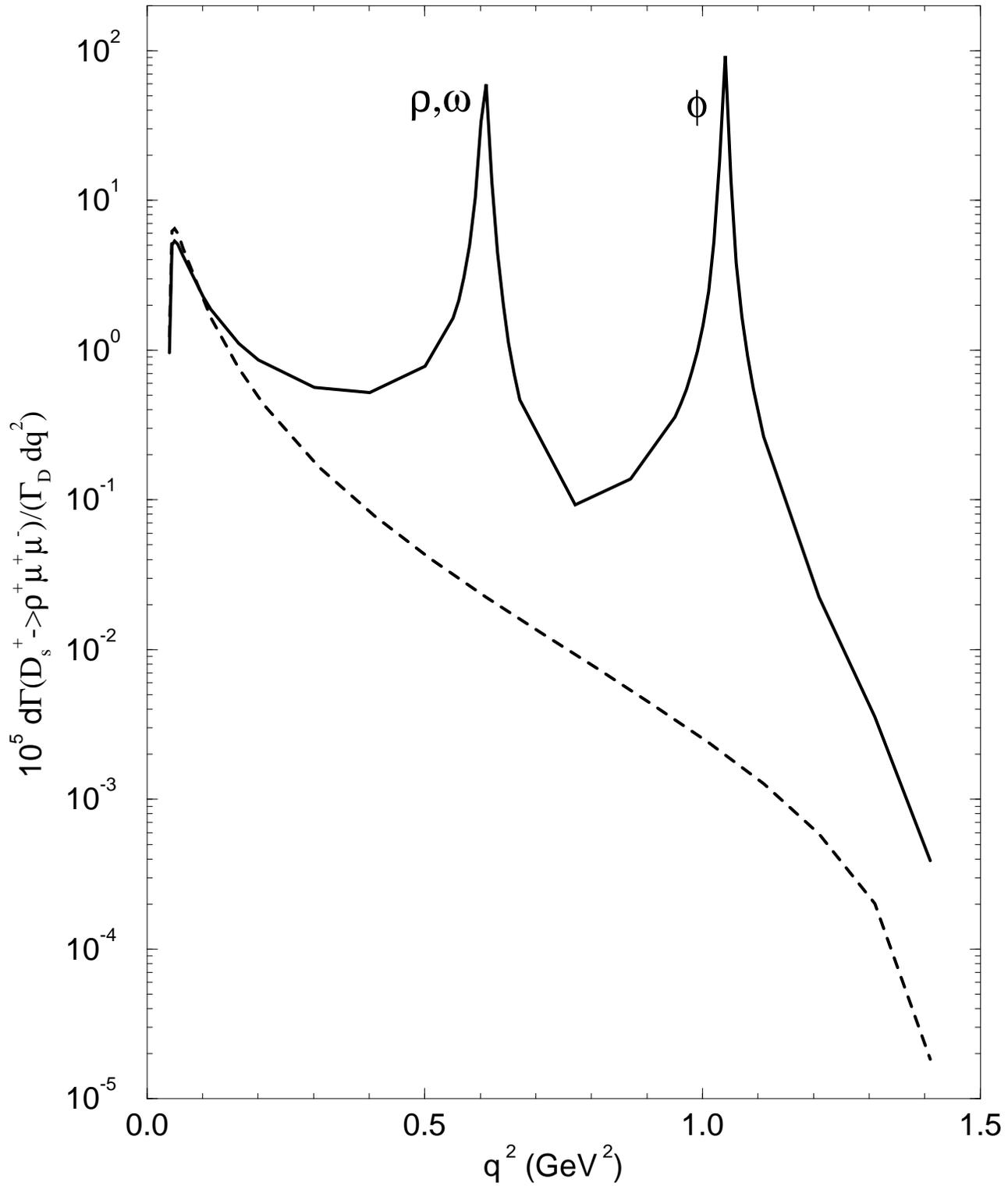
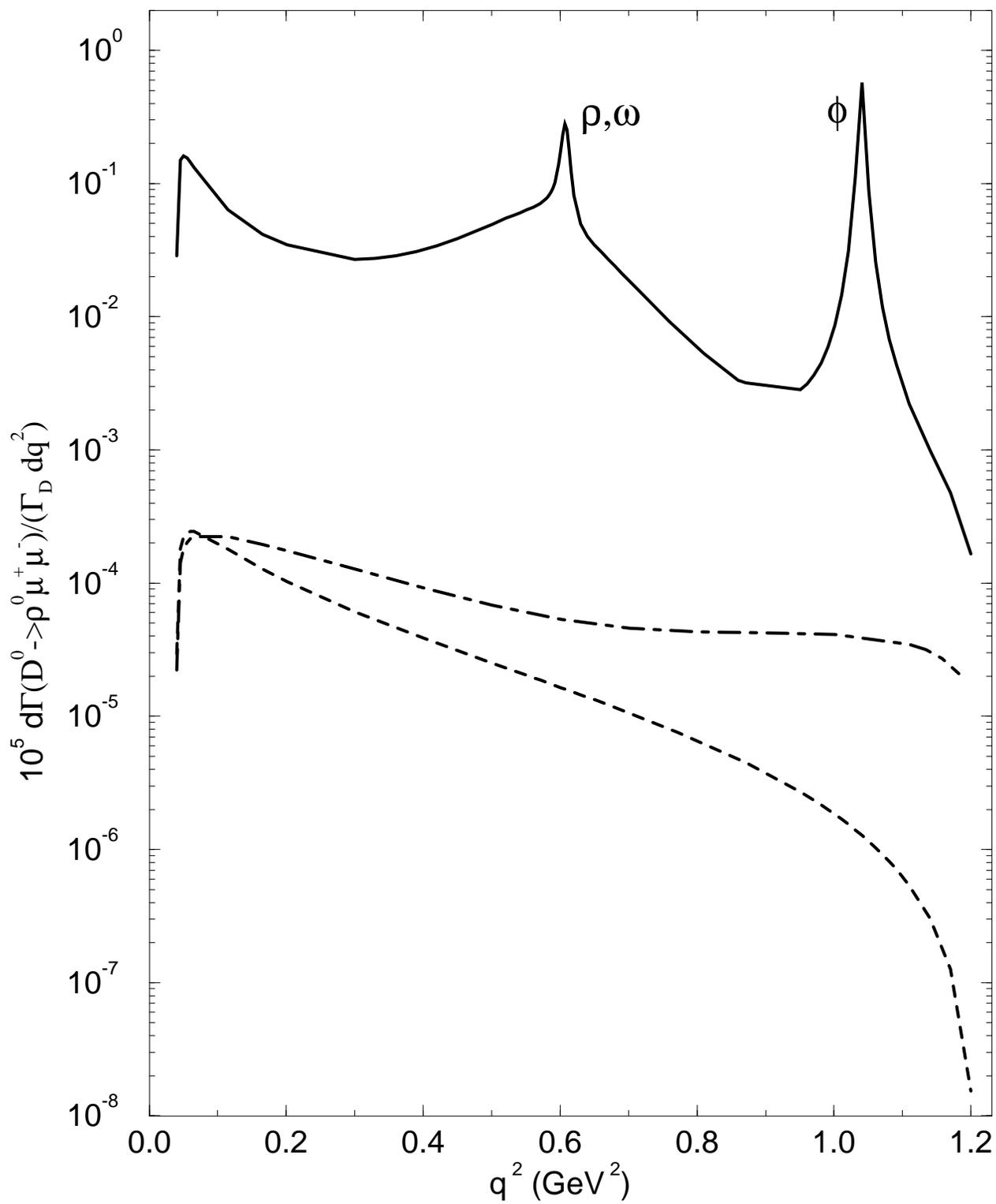
\end{document}